\documentclass[12pt,reqno]{amsart}
\allowdisplaybreaks[4]
\usepackage{graphicx} 
\usepackage{amssymb}
\usepackage{color}
\usepackage{amsmath}
\usepackage{cite}
\usepackage{subfigure}
\usepackage{graphicx}
\usepackage{epstopdf}
\usepackage{bibentry}
\usepackage{ulem}

\newtheorem{theorem}{Theorem}

\begin{document}

\title{Theoretical and experimental evidence of non-symmetric doubly localized rogue waves}

\author{Jingsong He$^{1,2\ast}$, Lijuan Guo$^{1}$, Yongshuai Zhang$^1$, and Amin Chabchoub$^3$}

\dedicatory {$^1$Department of Mathematics, Ningbo University, Ningbo, Zhejiang 315211, China\\
 $^2$ DAMTP, University of Cambridge, Cambridge, CB3 0WA, UK\\
$^3$ Centre for Ocean Engineering Science and Technology, Swinburne University of Technology, Hawthorn, Victoria 3122, Australia}

\thanks{$^*$ Corresponding author: hejingsong@nbu.edu.cn, jshe@ustc.edu.cn}

\begin{abstract}
We present determinant expressions for vector rogue wave solutions of the Manakov system, a two-component coupled nonlinear Schr\"odinger equation. As special case, we generate a family of exact and non-symmetric rogue wave solutions of the nonlinear Schr\"odinger equation up to third-order, localized in both space and time. The derived non-symmetric doubly-localized second-order solution is generated experimentally in a water wave flume for deep-water conditions. Experimental results, confirming the characteristic non-symmetric pattern of the solution, are in very good agreement with theory as well as with numerical simulations, based on the modified nonlinear Schr\"odinger equation, known to model accurately the dynamics of weakly nonlinear wave packets in deep-water.
\end{abstract}

\maketitle
%

\section{Introduction}
Rogue waves (RWs) appear in variety of physical systems, such as in oceans \cite{Kharif1,Kharif2,Osborne}, in atmosphere \cite{JPP75}, in optics \cite{Natural,SR2463}, in plasma \cite{EL96,Sharma1}. A fundamental property of such extreme waves is that they appear from nowhere, reach heights up to twice the height of surrounding waves, and disappear without a trace \cite{PLA373}. Significant theoretical progress has been achieved in recent years in order to describe the formation of RWs within the framework in weakly nonlinear evolution equations \cite{Osborne,ZakharovGelash}. In fact, RWs may be modeled in a general context by exact breather solutions of the nonlinear Schr\"odinger equation (NLS) \cite{Physrep52847}, which describe the propagation of stationary and pulsating wave packet dynamics in nonlinear dispersive media. An appropriate model to describe strong localizations in the medium of interest is the family of doubly-localized breather solutions, also referred to as Akhmediev-Peregrine (AP) solutions. The first-order solution is known as the Peregrine breather (PB) solution \cite{JAMSSB}; it is localized in space and in time, since it is the limiting case of time-periodic Kuznetsov-Ma solitons \cite{Kuznetsov,Ma} as well as space-periodic Akhmediev breathers \cite{AEK1985,tmp691089}, when the period of these periodic solutions tends to infinity. Furthermore, the AP amplifies the amplitudes of the carrier by a factor of three and higher. PB dynamics has been validated experimentally in optics \cite{NP6,ol36112}, in water waves \cite{PRL106,PRE86016311} and in plasma \cite{PRL10725}. More recently higher-order breathers have been observed as well \cite{PRX2011015,PRE86056601,Chabchoub4,FrisquetPRX,FrisquetPRA}.

There are indeed several interesting RW pattern solutions which the NLS may admit. Fundamental patterns may consist of a
simple central high peak, surrounded by several gradually decreasing solitonic peaks, an equal height triangular pattern, or a
circular pattern \cite{AkhmedievCascade}. Doubly-localized AP solutions are symmetric with respect to both the spatial and the temporal co-ordinate in theory. However, several  non-symmetrical rogue waves under fundamental pattern  have been observed in  water tank, see for instance
Figs. 4-6 in \cite{PRE86016311}, Figs. 3, 4 and 6 in  \cite{PRX2011015}, Figs. 3-9 in \cite{PRE86056601}. The observed non-symmetry of wave profiles in the wave flume is due to higher-dispersive and to the effect of the mean flow, not captured in the NLS, which can be explained by more accurate evolution equation, such as the generalized NLS in optics and the modified NLS (MNLS) in water waves \cite{Dysthe,DystheTrulsen,ChabchoubSupercontinuum}.

Independently from these higher-order effects, it is an interesting task to derive doubly-localized and non-symmetric solutions of the NLS. Physically, non-symmetric structures have different and fundamental distribution of energy and would be of interest for several physical applications. In this work, we derive exact solutions of the Manakov system up to third-order and as a special class  corresponding family of non-symmetric doubly localized solutions of the NLS. In addition, we demonstrate the physical accuracy of the derived solutions to describe strong localizations in water waves by generating experimentally the derived second-order solution in a water wave tank. It is shown that the expected amplitude amplification of the background is reached. Moreover, the experimental results are in very good agreement with numerical MNLS simulations.

This paper is organized as follows. In section 2, we introduce the Lax pair of the Manakov system and solve the Lax pair equations, with spectral parameter $\lambda$, from a periodic \textit{seed} solution.  In other words, we derive eigenfunctions of the Manakov
system, associated with the spectral parameter $\lambda$. In section 3, we give a determinant representation of the $n$th-order vector RWs of the Manakov system. In particular, exact expressions for the first-, second- and third-order RW solutions of the NLS equation, localized in both time and space are reported. The latter solutions have the property to be non-symmetric with respect to the spatial co-ordinate. Furthermore, the differences between the non-symmetric and corresponding symmetric doubly-localized RWs of the NLS equation are discussed in detail. In section 4, we shall compare experimental data, related to the evolution of the non-symmetric solution, derived in this work, in a water wave tank, with NLS maximal wave profile predictions as well as with numerical MNLS simulations. Finally, we summarize the main results in section 5.

\section{Vector eigenfunctions for the Manakov system}

 Recently, RW solutions of the multi-component soliton equations have been derived. These solutions attracted the scientific interest \cite{EPJ,CPL28,PRE86036601,PRL109,PRE87,PRE88052914, N14, PRE87032910,jpsj81033002,pre86066603,pre87012913, pre88062925}.  The Manakov system \cite{SPJETP}
 is an important integrable two-component NLS system and can be applied to optical as well as hydrodynamic systems \cite{OnoratoOsorneSerio,GramstadTrulsen}.
 In particular, if $q_1=\alpha q_2(\alpha \not=0 \in \mathbb{R})$ is a solution of the  Manakov system
  \begin{equation}\label{CNLS}
\begin{aligned}
{\rm i}q_{1t}+q_{1xx}+2(|q_1|^2+|q_2|^2)q_1=0,\\
{\rm i}q_{2t}+q_{2xx}+2(|q_2|^2+|q_1|^2)q_2=0, \\
\end{aligned}
\end{equation}
then
\begin{equation}\label{transformationnlsandcnls}
q=\sqrt{1+\frac{1}{\alpha^2}}q_1=\sqrt{1+\alpha^2}q_2
\end{equation}
is a solution of the NLS equation
\begin{equation}\label{NLS}
iq_t+q_{xx}+2q|q|^2=0.
\end{equation}
In the present paper, we shall construct the solutions of the Manakov system under the condition $q_1=\alpha q_2$ by the Darboux transformation (DT) from a periodic \textit{seed} solution, which imply several non-symmetrical extreme localizations of the NLS equation.

The DT is a powerful method to generate soliton \cite{AkhmedievBook,pla467100,actaamathsinica3143,Matveevbook,Gubook2,scichinaa121867,JPA40degasperis,JPA42degasperis},
breather \cite{actaamathsinica281713}, RW \cite{PRE80026601,PRE85026607,PRE87052914} solutions  of the NLS equation.
The breather solution of the Manakov system \cite{PhyicaD141104} has been constructed by the DT from a periodic \textit{seed}
 solution.
A crucial step of the DT method is to find the solutions, i.e. eigenfunctions of the linear partial differential equations,
defined by the Lax pair of the soliton equations. Therefore, we shall first solve the eigenfunctions of the Manakov system in this section in order to provide necessary preparatory work for the DT in next section.

The Manakov system \eqref{CNLS} is produced by the compatibility of the associated Lax pair equations:
\begin{equation}\label{laxpair}
\left\{
\begin{aligned}
\Psi_{x}&=U\Psi=({\rm i}J\lambda+Q)\Psi,\\
\Psi_{t}&=V\Psi=(2{\rm i}J\lambda^2+2Q\lambda+{\rm i}J(Q^2-Q_{x}))\Psi,
\end{aligned}\right.
\end{equation}
where
\begin{equation}
   J=\left(\begin{array}{ccc}
       -1 &0 &0\\
       0 &1 &0\\
       0 &0 &1\\
       \end{array}\right),\nonumber\\
       \quad Q=\left(\begin{array}{ccc}
       0 &q_1 &q_2\\
       -q_1^{*} &0 &0\\
       -q_2^{*} &0 &0\\
       \end{array}\right).
\end{equation}
Here, $\Psi$ is a vector eigenfunction associated with the spectral parameter $\lambda$.
\\

To obtain RW solutions of the Manakov system, we start with  periodic \textit{seed} solutions,
\begin{equation}\label{seed}
\begin{aligned}
q_1=c_1e^{{\rm i}(a_1x+b_1t)},\quad
q_2=c_2e^{{\rm i}(a_2x+b_2t)},
\end{aligned}
\end{equation}
where $a_1, a_2, b_1, b_2,c_1,c_2 \in \mathbb{R}$. Then, the dispersion relations can be written as
\begin{equation}\label{dispersion}
\begin{aligned}
b_1=2(c_1^2+c_2^2-a_1^2),\quad
b_2=2(c_1^2+c_2^2-a_2^2).
\end{aligned}
\end{equation}
In this work, we consider only a special case for the condition: $a_1=a_2=a$. Then, $b_1=b_2=b$, and
$\alpha=\frac{c_1}{c_2}$. The solutions $q_1$ and $q_2$ are therefore reduced to a solution of the NLS equation
by the straight-forward transformation \eqref{transformationnlsandcnls}. Hence, by n-fold DT,  we can generate new solutions
$q_1^{[n]}$ and  $q_2^{[n]}$ of the Manakov system from these \textit{seed} solutions, which satisfy the condition $q^{[n]}_1= \alpha q^{[n]}_2$, and then they are automatically reduced to new solutions $q^{[n]}$ of the NLS.

Now, we solve the eigenfunctions of the Lax pair, associated with \textit{seed} solutions \eqref{seed}. Firstly, we construct the following transformation $\Psi=R\Phi$ to map the variable coefficient partial differential equations \eqref{laxpair} to constant coefficient partial differential equations
 \begin{equation}\label{constantlaxpaireq}
 \Phi_{x}=\Omega\Phi, \ \ \Phi_t=\Lambda\Phi,
\end{equation}
by the following matrix
\begin{equation}\label{mapmattix}
R=\left( \begin{array}{ccc}
e^{\frac{{\rm i}(ax+bt)}{2}} &0 &0\\
0 &e^{\frac{-{\rm i}(ax+bt)}{2}} &0\\
0 &0 &e^{\frac{-{\rm i}(ax+bt)}{2}}\\
\end{array}\right).
\end{equation}
Here,
\begin{equation}\label{xpart}
\Omega=\left(\begin{array}{ccc}
-{\rm i}\lambda-{\rm i}\frac{a}{2} &c_1 &c_2\\
-c_1 &{\rm i}\lambda+{\rm i}\frac{a}{2} &0\\
-c_2 &0 &{\rm i}\lambda+{\rm i}\frac{a}{2} \\
\end{array}\right),
\end{equation}
\begin{equation}\label{tpart}
\Lambda=\left(\begin{array}{ccc}
-2{\rm i}\lambda^2+{\rm i}(c_1^2+c_2^2)-{\rm i}\frac{b}{2} &(2\lambda-a)c_1 &(2\lambda-a)c_2\\
(a-2\lambda)c_1 &2{\rm i}\lambda^2-{\rm i}c_1^2+{\rm i}\frac{b}{2} &-{\rm i}c_1c_2\\
(a-2\lambda)c_2 &-{\rm i}c_1c_2 &2{\rm i}\lambda^2-{\rm i}c_2^2+{\rm i}\frac{b}{2}\\
\end{array}\right),
\end{equation}
$\Phi(x,t,\lambda)=\left(\begin{matrix}
  f(x,t,\lambda),&g(x,t,\lambda),&h=(x,t,\lambda)
\end{matrix}\right)^T$, $\left(\cdot\right)^T$ denotes the transposition operator. Secondly, taking into account the dispersion relation \eqref{dispersion} in Eqs. \eqref{constantlaxpaireq}, \eqref{xpart} and \eqref{tpart},
then
\begin{equation}\label{efun}
\begin{aligned}
f=&\sum_{j=1}^{j=3}K_j{\rm e}^{u_jx+(2\lambda-a)u_jt},\\
g=&\sum_{j=1}^{j=3}K_j(\frac{-c_1}{u_j-{\rm i}\lambda-{\rm i}\frac{a}{2}}){\rm e}^{u_jx+(2\lambda-a)u_jt},\\
h=&\sum_{j=1}^{j=3}K_j(\frac{-c_2}{u_j-{\rm i}\lambda-{\rm i}\frac{a}{2}}){\rm e}^{u_jx+(2\lambda-a)u_jt},
\end{aligned}
\end{equation}
are solved by the method of characteristic equation. Here, $K_j(j=1,2,3)$ are three constants,
 $u_1$,$u_2$ and $u_3$ are three roots of the following equation
\begin{equation}\label{eigenfun}
\begin{aligned}
&{u}^{3}-\left(\frac{1}{2}\,{\rm i}a+{\rm i}\lambda \right) {u}^{2}+ \left( \frac{1}{4}\,{a}^{2}
+a\lambda+{c_{{2}}}^{2}+{\lambda}^{2}+{c_{{1}}}^{2} \right) u +\Gamma=0,
\end{aligned}
\end{equation}
with
$$\Gamma={\rm i}{
\lambda}^{3}+\frac{3}{2}\,{\rm i}a{\lambda}^{2}
+{\rm i}{c_{{1}}}^{2}\lambda+{\rm i}{c_{{2}}}
^{2}\lambda+\frac{3}{4}\,{\rm i}\lambda\,{a}^{2}+\frac{1}{2}\,{\rm i}{c_{{2}}}^{2}a+\frac{1}{8}\,{\rm i}{a}^{3}+\frac{1}{2}\,{\rm i}{c_{{1}}}^{2}a.$$
Finally, solving  Eq. \eqref{eigenfun}, we have
\begin{equation}\label{algebraroot}
\begin{aligned}
&u_1=\frac{\sqrt{-4\lambda^2-4\lambda a-a^2-4c_1^2-4c_2^2}}{2},\\
&u_2=\frac{-\sqrt{-4\lambda^2-4\lambda a-a^2-4c_1^2-4c_2^2}}{2},\\
&u_3={\rm i}(\frac{a}{2}+\lambda).
\end{aligned}
\end{equation}
Thus, eigenfunctions of the Manakov system associated with \textit{seed} Eq.\eqref{seed} are obtained by taking Eq.\eqref{algebraroot}, Eq.\eqref{efun} and  Eq.\eqref{mapmattix}
back into $\Psi=R\Phi$.
\\
\section{Rogue wave solutions of the Manakov system and the NLS}

In this section, we discuss the solutions of the Manakov system and the NLS equation.
Based on early work, where RW solutions have been derived \cite{jpsj81033002,pre86066603,pre87012913, pre88062925,PRE87052914,PhyicaD141104,JPA44305203,JMP53063507,cnsns191706,ps2014},  we set  $\lambda=\lambda_0+\epsilon^2=-\frac{a}{2}+{\rm i}\sqrt{c_1^2+c_2^2}+\epsilon^2$. Note that  $u_1=u_2=0$ and $u_3=-\sqrt{c_1^2+c_2^2}$ if $ \lambda=\lambda_0$. Furthermore, we set $K_1=1$, $K_2=-1$ and $K_3=0$, it is trivial to verify that $\Psi(\lambda_0)=0$. Thus, the Taylor expansion with respect to $\epsilon^2$ of the new solutions $q_1^{[n]}$ and $q_2^{[n]}$ generates the vector RWs of the Manakov system. For convenience, we introduce
$\Phi_j= \Phi(x,t,\lambda)|_{\lambda=\lambda_j}=\left(\begin{matrix}
  f_j,&g_j,&h_j
\end{matrix}\right)^T$, associated with $n$ distinct eigenvalues $\lambda_j, j=1, 2, 3, \cdots, n$.

\subsection{First-order rogue wave solutions of Manakov system}

We refer to \cite{PhyicaD141104} for the use of the DT to solve the Manakov system and we shall not repeat this procedure in this work.
As mentioned earlier, we set  $K_1=1$, $K_2=-1$, $K_3=0$ and $\lambda=\lambda_0+\epsilon^2$ in $q_1^{[1]}$ and $q_2^{[1]}$, then
 first-order Taylor expansion of them gives the first-order RW of the Manakov system.
\begin{theorem}\label{thm_1rw}
Let $\Psi_{1}=R\Phi_1$ be a solution  of the Lax pair equations associated with $\lambda_{1}$ and \textit{seed} solution,  $\lambda_1=\lambda_0+\epsilon^2$, then $(q_1^{[1]},q_2^{[1]})$ given by the following
 determinant representation are the first-order vector RW of the Manakov system:
 \begin{equation}\label{f_1_rw}
 q_1^{[1]}=\left(c_1+2{\rm i}\frac{|\Omega_{1}'|}{|\Omega_{2}'|}\right){\rm e}^{{\rm i}(ax+bt)},\quad q_2^{[1]}=\left(c_2-2{\rm i}\frac{|\Omega_{3}'|}{|\Omega_{2}'|}\right){\rm e}^{{\rm i}(ax+bt)},
 \end{equation}
 where
 \begin{equation*}
 \begin{aligned}
 \Omega_{1}=\left(\begin{matrix}
 f_{1} &h_{1} &\lambda_1 f_{1} \\
 -g_{1}^* &0 &-\lambda_1^*g_{1}^*\\
 -h_{1}^* &f_{1}^* &-\lambda_1^*h_{1}^*\\
 \end{matrix}\right),\quad
 \Omega_{2}=\left(\begin{matrix}
 f_{1} &g_{1} &h_{1} \\
 -g_{1}^* &f_{1}^* &0\\
 -h_{1}^* &0 &f_{1}^*\\
 \end{matrix}\right),\quad
 \Omega_{3}=\left(\begin{matrix}
 f_{1} &g_{1} &\lambda_1 f_{1} \\
 -g_{1}^* &f_{1}^* &-\lambda_1^*g_{1}^*\\
 -h_{1}^* &0 &-\lambda_1^*h_{1}^*\\
 \end{matrix}\right),
 \end{aligned}
\end{equation*}
and
\begin{equation}
\begin{aligned}
  \Omega_{1}'=&\left(\left.\frac{\partial^{2}}{\partial{\epsilon^{2}}}\right|_{\epsilon=0}\left(\Omega_{1}\right)
  _{ij}\left(\lambda_0+\epsilon^2\right)\right)_{3\times3},\\
  \Omega_{2}'=&\left(\left.\frac{\partial^{2}}{\partial{\epsilon^{2}}}\right|_{\epsilon=0}\left(\Omega_{2}\right)
  _{ij}\left(\lambda_0+\epsilon^2\right)\right)_{3\times3},\\
  \Omega_{3}'=&\left(\left.\frac{\partial^{2}}{\partial{\epsilon^{2}}}\right|_{\epsilon=0}\left(\Omega_{3}\right)
  _{ij}\left(\lambda_0+\epsilon^2\right)\right)_{3\times3}.
\end{aligned}
\end{equation}
\end{theorem}

\begin{figure}[h]
\subfigure[]{\includegraphics[width=6.5cm]{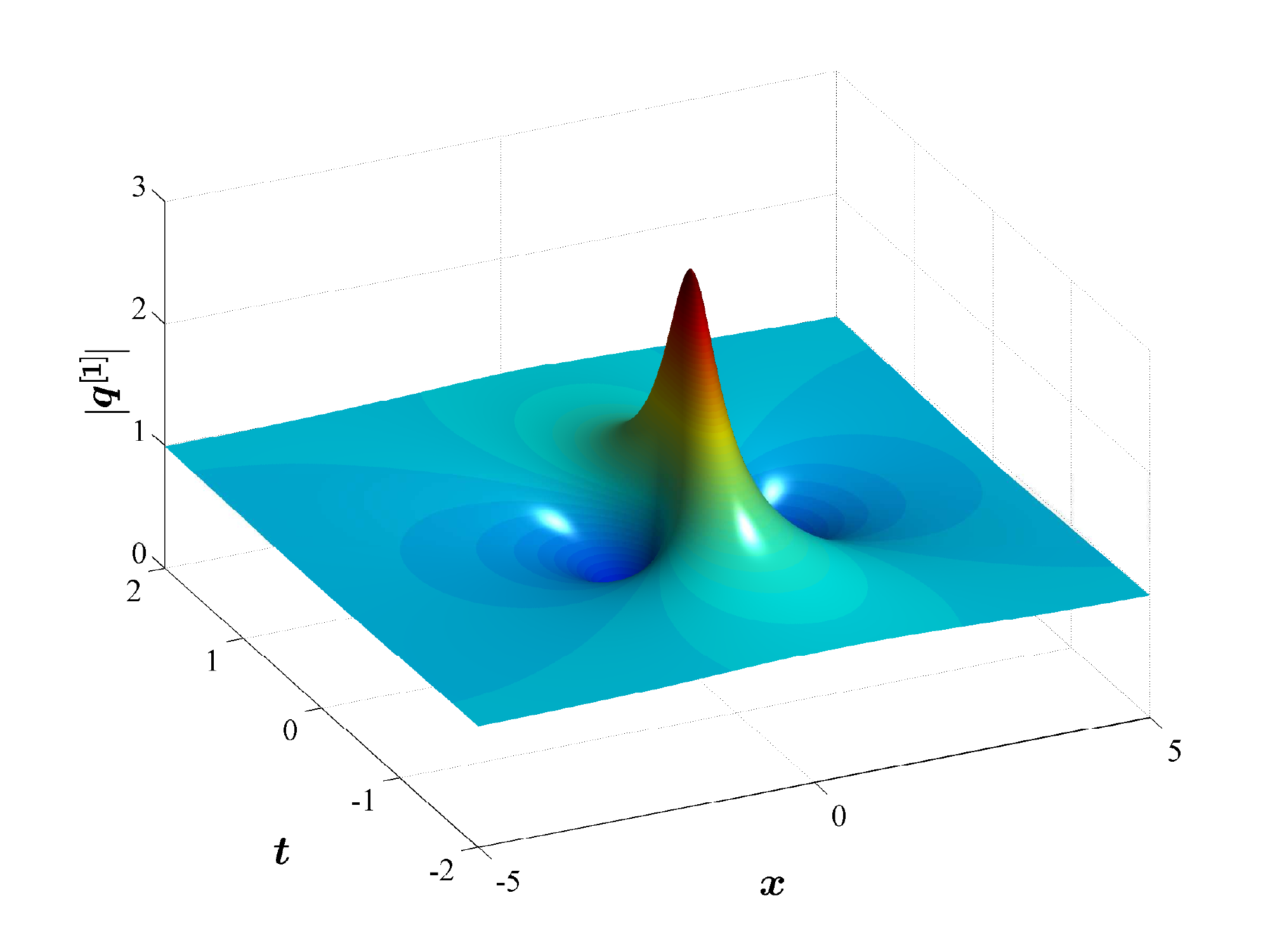}}
\subfigure[]{\includegraphics[width=6.5cm]{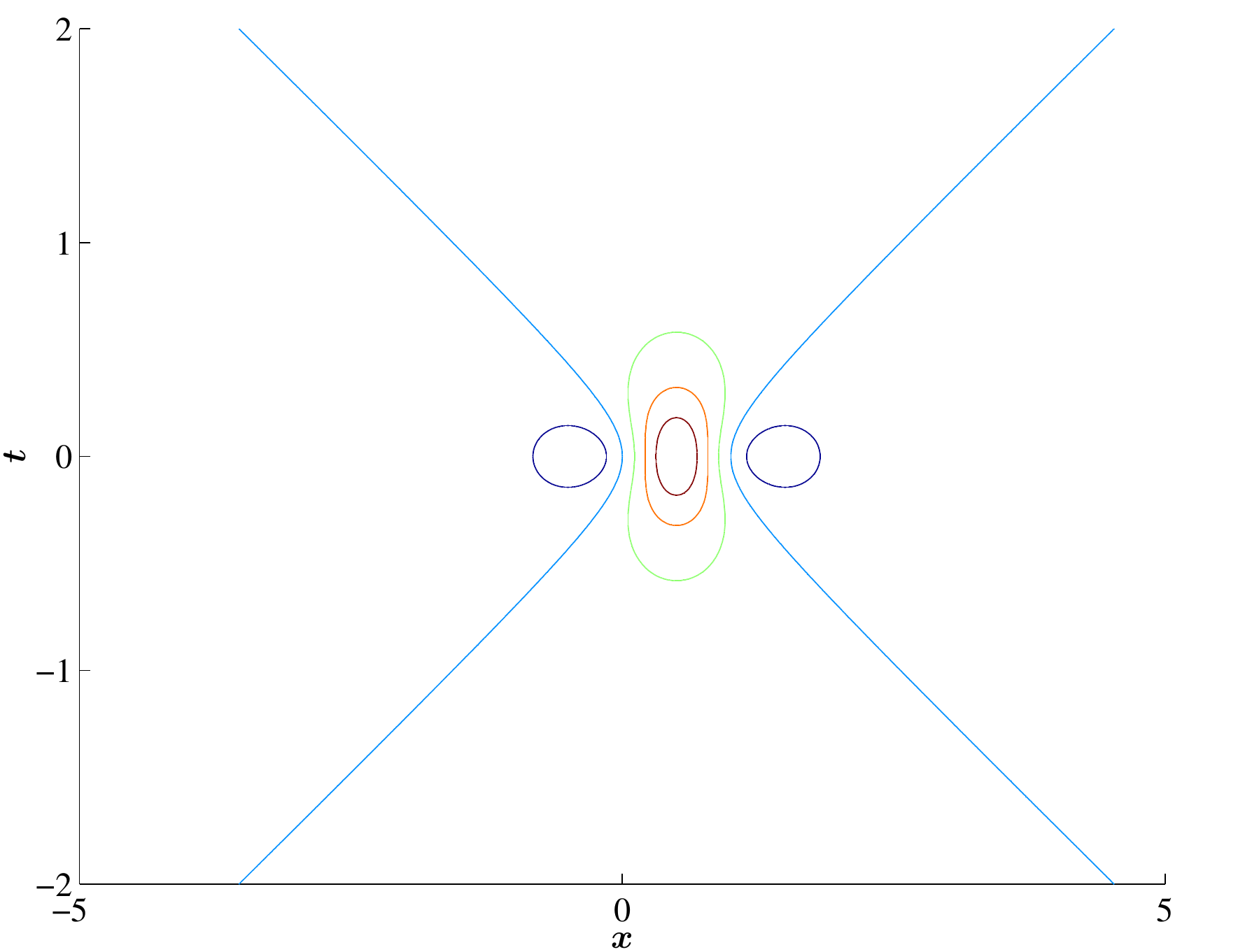}}
\caption{(a) Evolution of the first-order doubly-localized RW non-symmetric $q^{[1]}$. (b) Contour plot of the solution $q^{[1]}$.}
\label{fig1}
\end{figure}

Substituting \eqref{efun} into \eqref{f_1_rw} of \textcolor{red}{T}heorem \ref{thm_1rw}, the explicit expressions of the the first-order vector RWs
of the Manakov system are given as
\begin{equation}\label{1_rw}
\begin{aligned}
q_1^{[1]}=e^{{\rm i}(ax+bt)}(c_1+\frac{F_1}{H_1}),\quad
q_2^{[1]}=e^{{\rm i}(ax+bt)}(c_2+\frac{G_1}{H_1}),
\end{aligned}
\end{equation}
where
\begin{equation}\nonumber
\begin{aligned}
&F_1=4c_{{1}}\,\sqrt {{c_{{1}}}^{2}+{c_{{2}}}^{2}}L_1L_2, \ \ \ \ \ \ \ \ \  G_1=4c_{{2}}\,\sqrt {{c_{{1}}}^{2}+{c_{{2}}}^{2}}L_1L_2,\\
&L_1=x-2\,ta+2\,{\rm i}t\sqrt {{c_{{1}}}^{2}+{c_{{2}}}^{2}},\ \  L_2=2\,{\rm i}t \left( {c_{{1}}}^{2}+{c_{{2}}}^{2} \right) + \left( 2\,t
a -\,x\right) \sqrt {{c_{{1}}}^{2}+{c_{{2}}}^{2}}+1,\\
&H_1=8\,{t}^{2} \left( {c_{{1}}}^{2}+{c_{{2}}}^{2} \right) ^{2}+ \left(2\,{x}^{2}-
8\,xta+8\,{t}^{2}{a}^{2}\right)  \left( {c_{{
1}}}^{2}+{c_{{2}}}^{2} \right) + \left(4\,ta-2\,x \right) \sqrt {{
c_{{1}}}^{2}+{c_{{2}}}^{2}}+1.\\
\end{aligned}
\end{equation}
It is obvious that ${q_1^{[1]}}/{q_2^{[1]}}={c_1}/{c_2}$, so that $q^{[1]}=\sqrt{1+\frac{c^2_2}{c^2_1}}q_1^{[1]}=\sqrt{1+\frac{c^2_1}{c^2_2}}q_2^{[1]}$
 is the first-order RW of the
NLS,  which is plotted in Fig. \ref{fig1}.

By choosing $c_1=c_2=\frac{\sqrt{2}}{2}$, the amplitude of the plane wave, i.e. carrier of $q^{[1]}$ is of one. The maximum amplitude of the RW $q^{[1]}$ is located at $(\frac{1}{2},0)$. Compared with the results of \cite{N14} (Eqs. 10(a) and 10(b), Figs. 1, 2 and 3), it can be concluded that $q^{[1]}$ possesses the same dynamics as the doubly-localized Peregrine solution \cite{JAMSSB}, except for the location of the rogue peak. 

\begin{figure}[h]
\subfigure[]{\includegraphics[width=6.5cm]{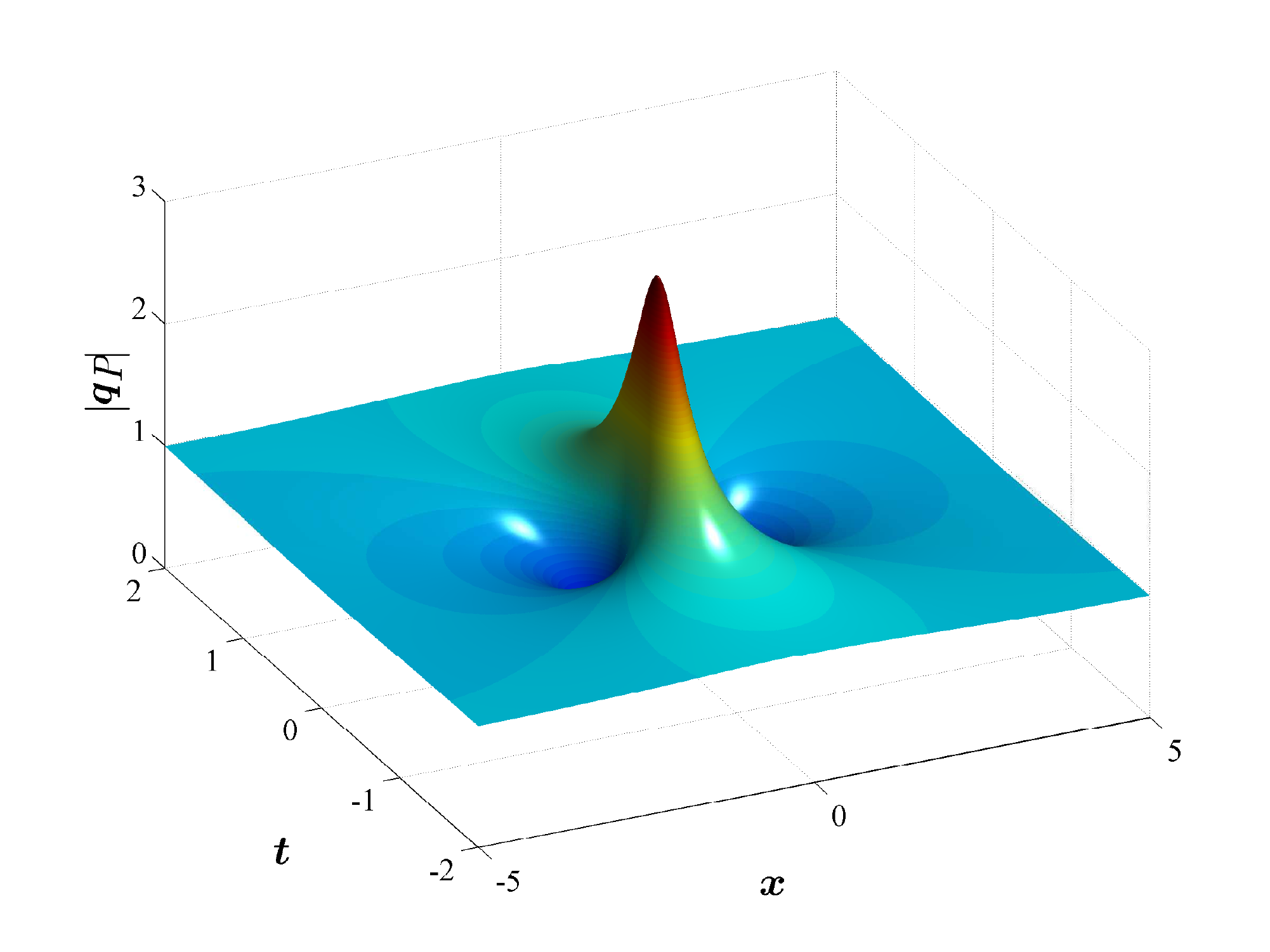}}
\subfigure[]{\includegraphics[width=6.5cm]{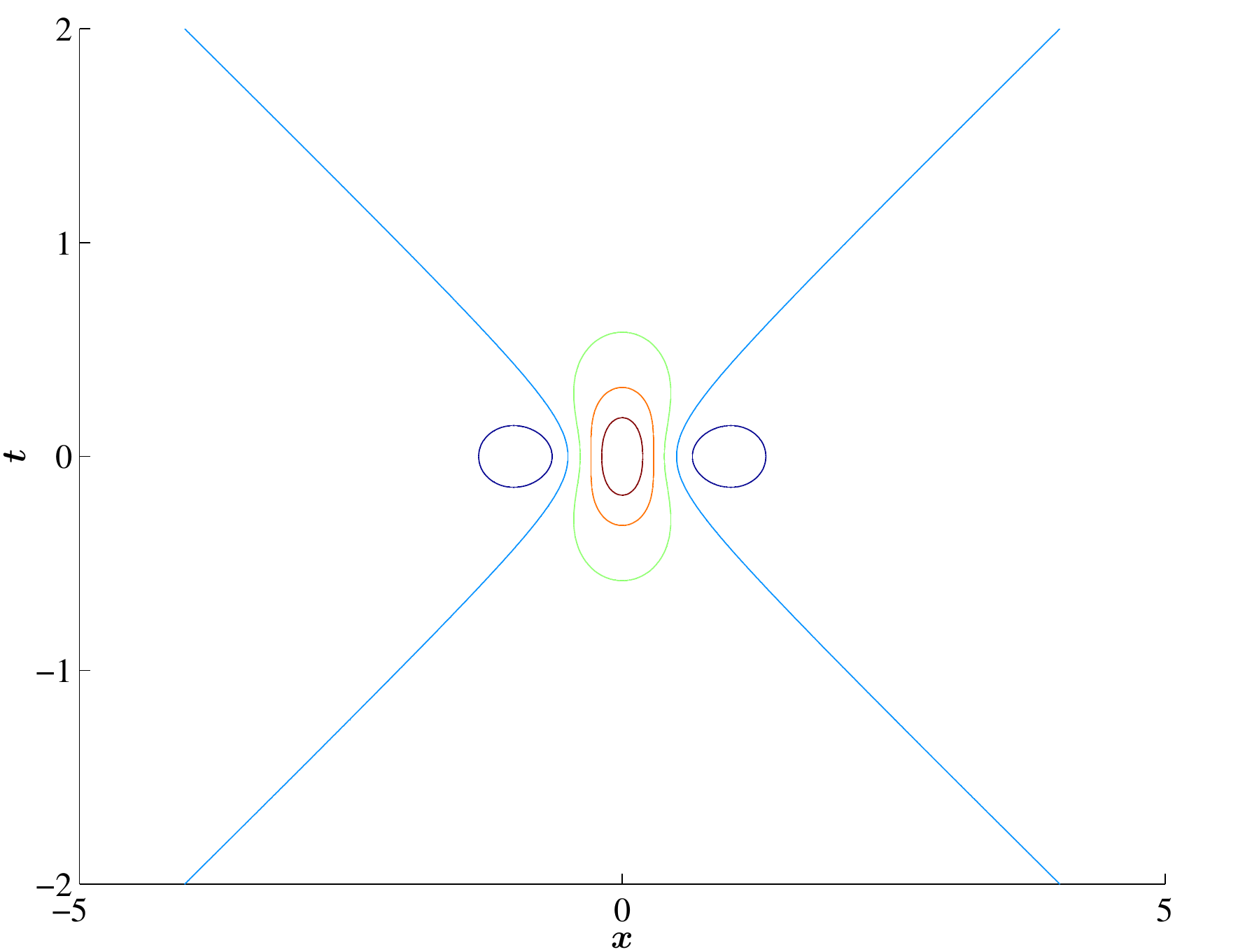}}
\caption{(a) Temporal and spatial evolution of the Peregrine breather. (b) Contour plot of the Peregrine breather.}
\label{fig2}
\end{figure}

Note that $q^{[1]}$  is non-symmetric with respect to ($x=0$)-axis, as displayed in Fig. \ref{fig1}. However, after applying a straight-forward transformation $x=x+\frac{1}{2}$, it becomes symmetric.
 Moreover, let $\,c_1=\frac{\sqrt{2}}{2}$ and $c_2=\frac{\sqrt{2}}{2}$ in this case, then $\alpha=1$, and
\begin{equation}
  q_s^{[1]}=\sqrt{2}q_1^{[1]}=\sqrt{2}q_2^{[1]}={\frac { \left( 3-16\,{t}^{2}+16\,{\rm i}t-4
\,{x}^{2} \right) }{1+16\,{t}^{2}+4\,{x}^{2}}}{{\rm e}^{2\,{\rm i}t}}
\end{equation}
is the Peregrine soliton of the NLS equation \cite{JAMSSB}, as expected, see Fig. \ref{fig2}.

This fact means that the asymmetry feature in the first-order RW  is non-essential, since it can be explained by a simple linear shift of the spatial variable. We need to study higher-order RWs of the NLS to explore the non-symmetric property of such solutions.

\subsection{Higher-order rogue wave solutions of the Manakov system}

\begin{figure}[h]
\subfigure[]{\includegraphics[width=6.5cm]{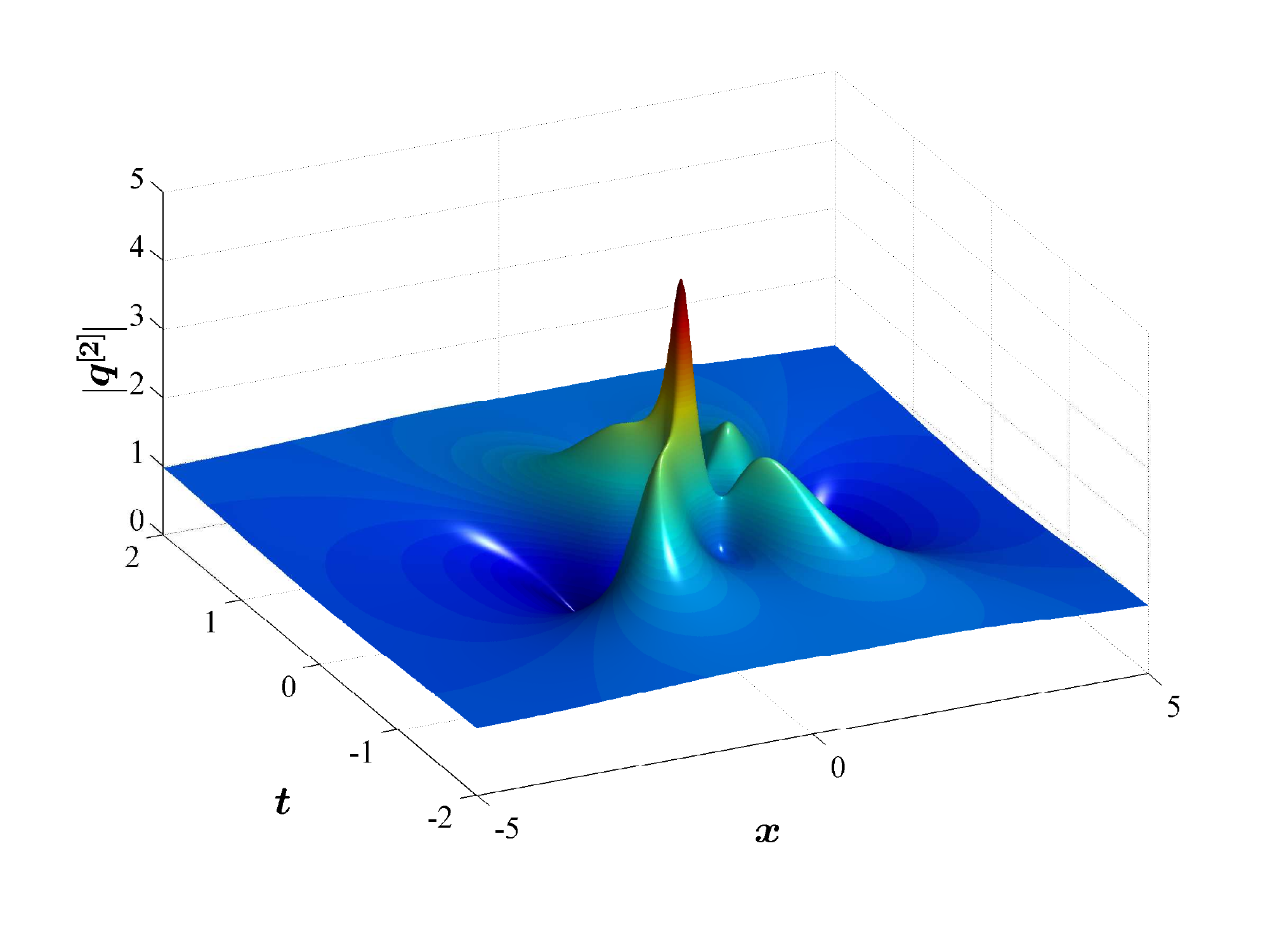}}
\subfigure[]{\includegraphics[width=6.5cm]{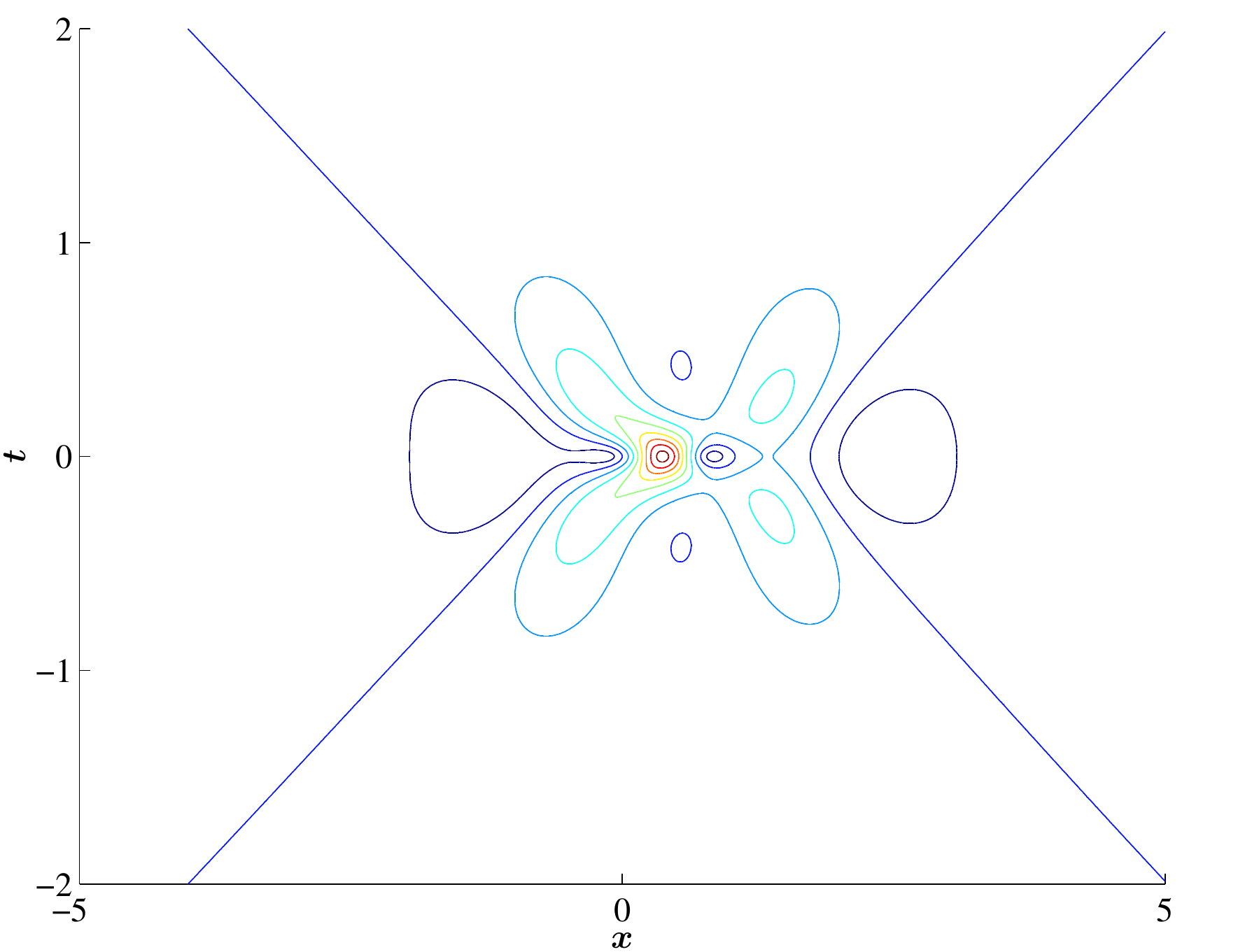}}
\caption{(a) Evolution of the second-order doubly-localized RW non-symmetric $q^{[2]}$. (b) Contour plot of the solution $q^{[2]}$.}
\label{fig3}
\end{figure}

Similar to the case of NLS \cite{scichinaa121867, PRE87052914} and according to the one-fold DT, we obtain the determinant expressions of $n$th-order solution $(q_1^{[n]},\,q_2^{[n]})$ through iteration. With the method of Taylor expansion with respect to $\epsilon^2$ in $(q_1^{[n]},\,q_2^{[n]})$ through $\lambda_j=\lambda_0+\epsilon^2$,
the determinant expressions of the $n$th-order RWs are obtained as well.
\begin{theorem}\label{thm_nrw}
Let $\Psi_{i}=R\Phi_i$ $(i=1,2,\cdots,n)$ be $n$ distinct eigenfunctions associated with  $\lambda_{i}(i=1,2,\cdots,n)$,
then, $(q_1^{[n]},\,q_2^{[n]})$ given by the following determinant representations are the $n$th-order RWs of the Manakov system.
\begin{equation}
q_1^{[n]}=(c_1+2{\rm i}\frac{|\Omega_{1}'|}{|\Omega_{2}'|}){\rm e}^{{\rm i}(ax+bt)},\quad q_2^{[n]}=(c_2-2{\rm i}\frac{|\Omega_{3}'|}{|\Omega_{2}'|}){\rm e}^{{\rm i}(ax+bt)},
\end{equation}
where
\begin{equation}
\nonumber
\Omega_{1}=\left(\begin{matrix}
f_{1} &g_{1} &h_{1} &\lambda_1f_{1} &\lambda_1g_{1} &\lambda_1h_{1} &\cdots
&\lambda_1^{n-1}f_{1} &\lambda_1^{n-1}h_{1} &\lambda_1^{n}f_{1}\\
-g_{1}^* &f_{1}^* &0 &-\lambda_1^*g_{1}^* &\lambda_1^*f_{1}^* &0 &\cdots
&-(\lambda_1^*)^{n-1}g_{1}^* &0 &-(\lambda_1^*)^{n}g_{1}^*\\
-h_{1}^* &0 &f_{1}^* &-\lambda_1^*h_{1}^* &0 &\lambda_1^*f_{1}^* &\cdots
&-(\lambda_1^*)^{n-1}h_{1}^* &(\lambda_1^*)^{n-1}f_{1}^* &-(\lambda_1^*)^{n}h_{1}^*\\
\vdots &\vdots &\vdots &\vdots &\vdots &\vdots &\vdots &\vdots &\vdots &\vdots\\
f_{n} &g_{n} &h_{n} &\lambda_nf_{n} &\lambda_ng_{n} &\lambda_nh_{n} &\cdots
&(\lambda_n^*)^{n-1}f_{n} &(\lambda_n^*)^{n-1}h_{n} &(\lambda_n^*)^{n}f_{n}\\
-g_{n}^* &f_{n}^* &0 &-\lambda_n^*g_{n}^* &\lambda_n^*f_{n}^* &0 &\cdots
&-(\lambda_n^*)^{n-1}g_{n}^* &0 &-(\lambda_n^*)^{n}g_{n}^*\\
-h_{n}^* &0 &f_{n}^* &-\lambda_n^*h_{n}^* &0 &\lambda_n^*f_{n}^* &\cdots
&-(\lambda_n^*)^{n-1}h_{n}^* &(\lambda_n^*)^{n-1}f_{n}^* &-(\lambda_n^*)^{n}h_{n}^*\\
\end{matrix}\right),\\
\end{equation}
\begin{equation}\nonumber
\begin{aligned}
\Omega_{2}=\left(\begin{matrix}
f_{1} &g_{1} &h_{1} &\lambda_1f_{1} &\lambda_1g_{1} &\lambda_1h_{1} &\cdots
&\lambda_1^{n-1}f_{1} &\lambda_1^{n-1}g_{1} &\lambda_1^{n-1}h_{1}\\
-g_{1}^* &f_{1}^* &0 &-\lambda_1^*g_{1}^* &\lambda_1^*f_{1}^* &0 &\cdots
&-(\lambda_1^*)^{n-1}g_{1}^* &(\lambda_1^*)^{n-1}f_{1}^* &0\\
-h_{1}^* &0 &f_{1}^* &-\lambda_1^*h_{1}^* &0 &\lambda_1^*f_{1}^* &\cdots
&-(\lambda_1^*)^{n-1}h_{1}^* &0 &(\lambda_1^*)^{n-1}f_{1}^*\\
\vdots &\vdots &\vdots &\vdots &\vdots &\vdots &\vdots &\vdots &\vdots &\vdots\\
f_{n} &g_{n} &h_{n} &\lambda_nf_{n} &\lambda_ng_{n} &\lambda_nh_{n} &\cdots
&\lambda_n^{n-1}f_{n} &\lambda_n^{n-1}g_{n} &\lambda_n^{n-1}h_{n}\\
-g_{n}^* &f_{n}^* &0 &-\lambda_n^*g_{n}^* &\lambda_n^*f_{n}^* &0 &\cdots
&-(\lambda_n^*)^{n-1}g_{n}^* &(\lambda_n^*)^{n-1}f_{n}^* &0\\
-h_{n}^* &0 &f_{n}^* &-\lambda_n^*h_{n}^* &0 &\lambda_n^*f_{n}^* &\cdots
&-(\lambda_n^*)^{n-1}h_{n}^* &0 &(\lambda_n^*)^{n-1}f_{n}^*\\
\end{matrix}\right),\\
\Omega_{3}=\left(\begin{matrix}
f_{1} &g_{1} &h_{1} &\lambda_1f_{1} &\lambda_1g_{1} &\lambda_1h_{1} &\cdots
&\lambda_1^{n-1}f_{1} &\lambda_1^{n-1}g_{1} &\lambda_1^{n}f_{1}\\
-g_{1}^* &f_{1}^* &0 &-\lambda_1^*g_{1}^* &\lambda_1^*f_{1}^* &0 &\cdots
&-(\lambda_1^*)^{n-1}g_{1}^* &(\lambda_1^*)^{n-1}f_{1}^* &-(\lambda_1^*)^{n}g_{1}^*\\
-h_{1}^* &0 &f_{1}^* &-\lambda_1^*h_{1}^* &0 &\lambda_1^*f_{1}^* &\cdots
&-(\lambda_1^*)^{n-1}h_{1}^* &0 &-(\lambda_1^*)^{n}h_{1}^*\\
\vdots &\vdots &\vdots &\vdots &\vdots &\vdots &\vdots &\vdots &\vdots &\vdots\\
f_{n} &g_{n} &h_{n} &\lambda_nf_{n} &\lambda_ng_{n} &\lambda_nh_{n} &\cdots
&(\lambda_n^*)^{n-1}f_{n} &(\lambda_n^*)^{n-1}g_{n} &(\lambda_n^*)^{n}f_{n}\\
-g_{n}^* &f_{n}^* &0 &-\lambda_n^*g_{n}^* &\lambda_n^*f_{n}^* &0 &\cdots
&-(\lambda_n^*)^{n-1}g_{n}^* &(\lambda_n^*)^{n-1}f_{n}^* &-(\lambda_n^*)^{n}g_{n}^*\\
-h_{n}^* &0 &f_{n}^* &-\lambda_n^*h_{n}^* &0 &\lambda_n^*f_{n}^* &\cdots
&-(\lambda_n^*)^{n-1}h_{n}^* &0 &-(\lambda_n^*)^{n}h_{n}^*\\
\end{matrix}\right),
\end{aligned}
\end{equation}
and
\begin{equation}
\begin{aligned}
  \Omega_{1}'=&\left(\left.\frac{\partial^{n_i}}{\partial{\epsilon^{n_i}}}\right|_{\epsilon=0}\left(\Omega_{1}\right)
  _{ij}\left(\lambda_0+\epsilon^2\right)\right)_{3n\times3n},\\
  \Omega_{2}'=&\left(\left.\frac{\partial^{n_i}}{\partial{\epsilon^{n_i}}}\right|_{\epsilon=0}\left(\Omega_{2}\right)
  _{ij}\left(\lambda_0+\epsilon^2\right)\right)_{3n\times3n},\\
  \Omega_{3}'=&\left(\left.\frac{\partial^{n_i}}{\partial{\epsilon^{n_i}}}\right|_{\epsilon=0}\left(\Omega_{3}\right)
  _{ij}\left(\lambda_0+\epsilon^2\right)\right)_{3n\times3n},
\end{aligned}
\end{equation}
$n_i=2[\frac{i-1}{3}]+2,\,[i]$ denotes the floor function of $i$.
\end{theorem}
For convenience, let $a=0,\,c_1=\frac{\sqrt{2}}{2}$ and $c_2=\frac{\sqrt{2}}{2}$ in the following. Then, $\alpha=1$, and $q^{[n]}=\sqrt{2}q_1^{[n]}=\sqrt{2}q_2^{[n]}$ is the $n$th-order RWs of the NLS equation.
Let $n=2$ in \textcolor{red}{T}heorem \ref{thm_nrw}, the second-order vector RWs of the Manakov system  are given by
\begin{equation}\label{2_rw}
\begin{aligned}
q_1^{[2]}=q_2^{[2]}=(\frac{\sqrt{2}}{2}+\frac{F_2}{H_2}){\rm e}^{2{\rm i}t},
\end{aligned}
\end{equation}
with
\begin{equation}\nonumber
\begin{aligned}
F_2=&12\,{\rm i}\sqrt {2} \left( 12\,t-32\,{t}^{3}-12\,xt+16\,{x}^{3}t+64\,{t}^{
3}x-8\,{x}^{4}t-64\,{x}^{2}{t}^{3}-128\,{t}^{5}\right.\\
&\left.+48\,{\rm i}{x}^{2}{t}^{2}-4
\,{\rm i}{x}^{3}+6\,{\rm i}{x}^{2}+160\,{\rm i}{t}^{4}-48\,{\rm i}x{t}^{2}+48\,{\rm i}{t}^{2}+2\,{\rm i}{x
}^{4}-3\,{\rm i}x \right),\\
H_2=&1024\,{t}^{6}+768\,{x}^{2}{t}^{4}-768\,{t}^{4}x+1920\,{t}^{4}
+192\,{x}^{4}{t}^{2}-384\,{x}^{3}{t}^{2}+288\,x{t}^{2}\\
&+288\,{t}^{2}+16
\,{x}^{6}-48\,{x}^{5}+72\,{x}^{4}-72\,{x}^{3}+72\,{x}^{2}-36\,x+9.
\end{aligned}
\end{equation}
It is a non-symmetrical doubly-localized RW $q^{[2]}=\sqrt{2}q_1^{[2]}$ of the NLS, and is displayed in Fig. \ref{fig3}.

It is obvious that the structure of this solution is highly non-symmetrical in the $(x,t)$ plane and the highest peak
is not located at $(0,0)$, unlike  the profile of the symmetrical doubly-localized AP breathers' fundamental pattern \cite{PRX2011015}. In fact, as can been observed qualitatively in Fig. \ref{fig3} (a), there are four small soliton peaks surrounding one main and highest peak. However, the left two soliton peaks with respect to the ($x=0$)-axis are higher than the two on the right. Besides, the highest peak is not located on the center of the four small peaks as the symmetrical fundamental pattern\cite{pla3732137}. It is located on the left side of the co-ordinate origin center. Specifically,  the co-ordinate of the main peak is $(0.372, 0)$ in the $(x,t)$ plane,  and the height of the main peak is of 4.695. This value corresponds to the amplitude amplification of this solution, since the carrier amplitude is of one. This also means that the amplitude of this RW  is not distributed symmetrically along $x$-axis.
 In Fig. \ref{fig3} (a), it can be also seen that there is only one significant trough at the left side of the main peak. This is not the case for the symmetric AP solution, where two identical and significant troughs can be noticed at each side of the main peak along ($t=0$)-axis, see \cite{pla3732137}. In order to depict these troughs, we have displayed the contour line at asymptotic plane in Fig. \ref{fig3} (b). Generally, Fig. \ref{fig3} (a) and (b) show the noticeable and clear differences to the dynamics, described by the symmetrical fundamental pattern of the second-order NLS RW \cite{pla3732137}. The centers of the depicted three circles are given by (0.6,0.41), (0.84, 0), (0.6,-0.41) in the ($x,t$)-plane.

\begin{figure}[h]
\subfigure[]{\includegraphics[width=6.5cm]{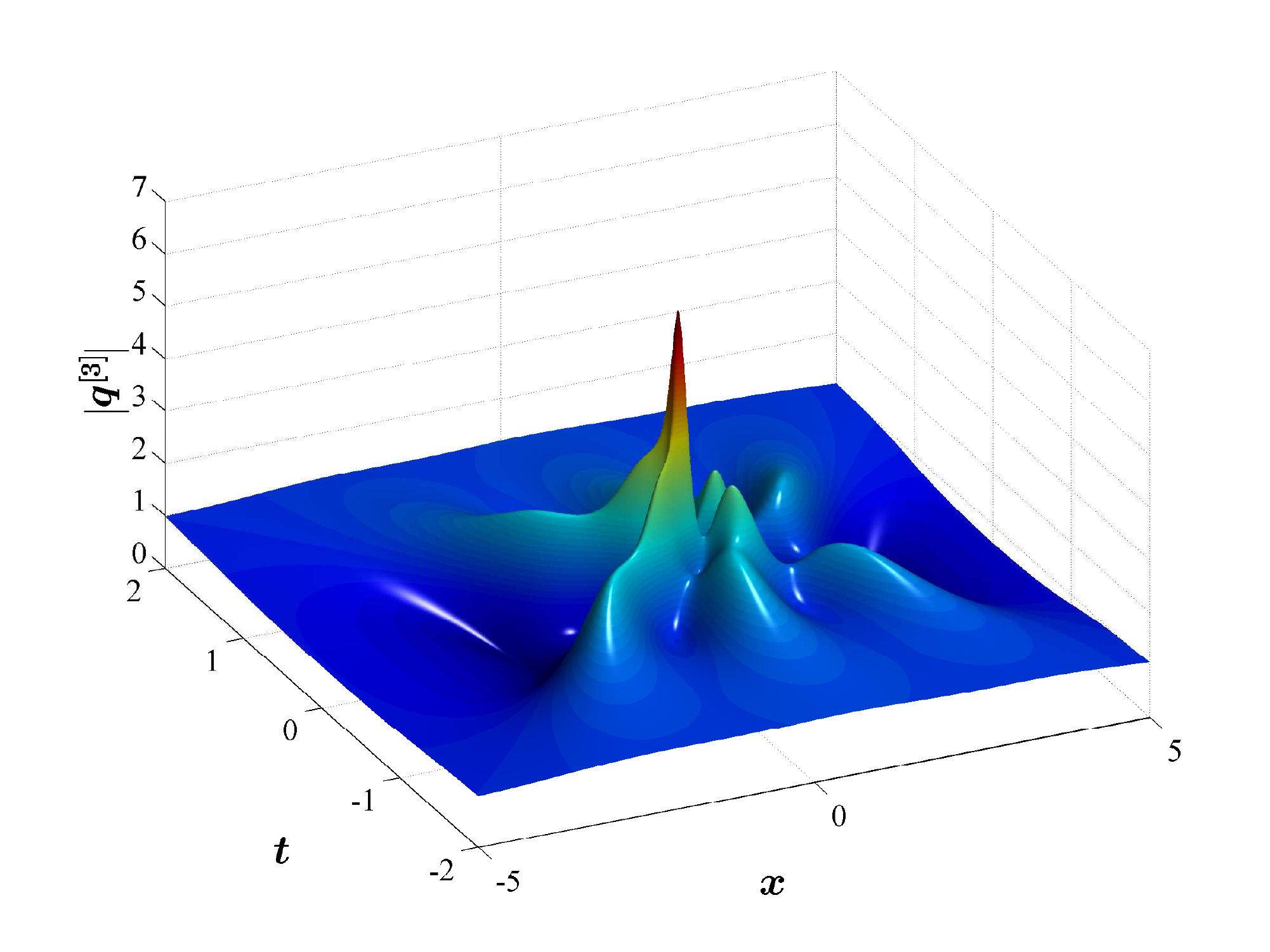}}
\subfigure[]{\includegraphics[width=6.5cm]{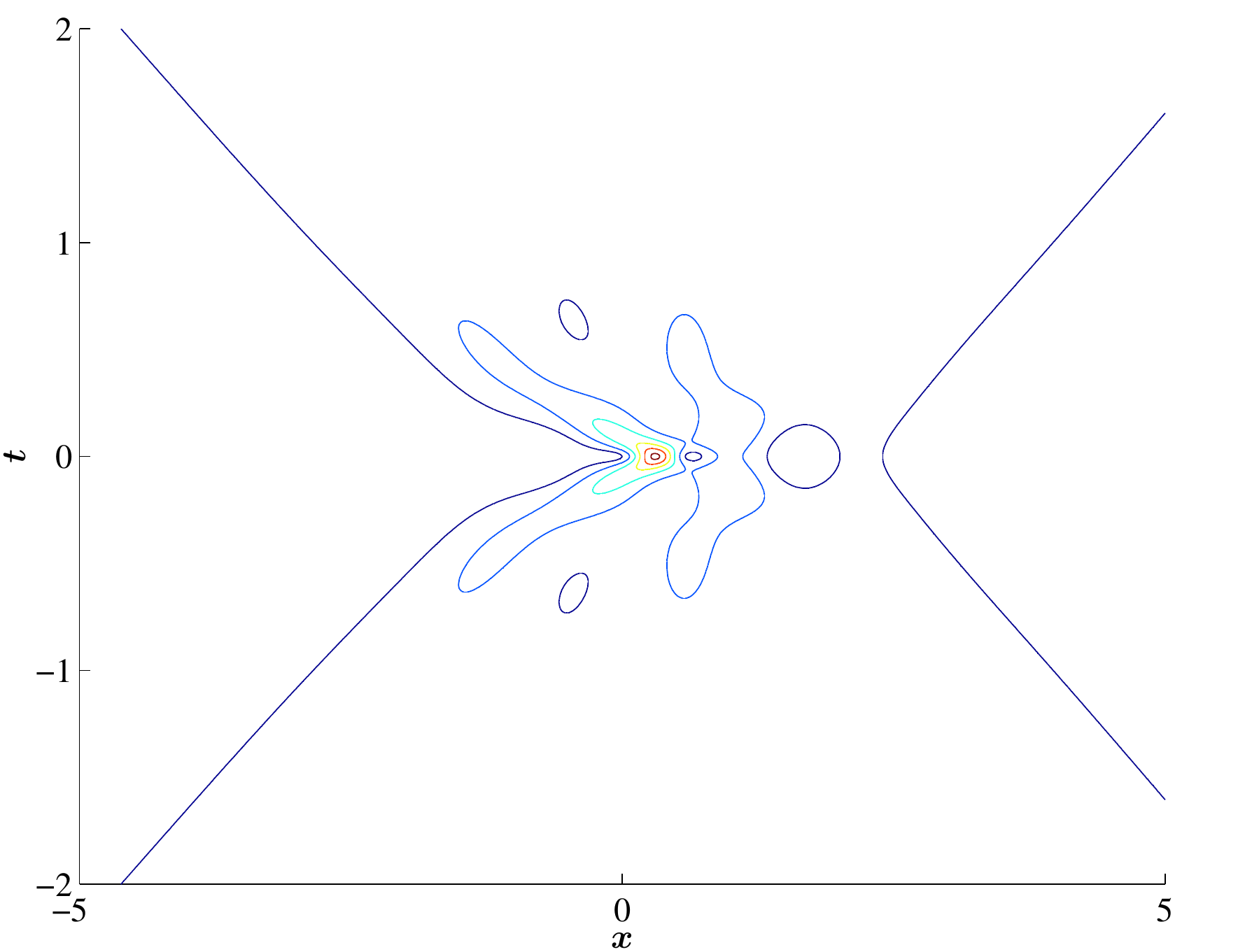}}
\caption{(a) Evolution of the third-order doubly-localized RW non-symmetric $q^{[3]}$. (b) Contour plot of the solution $q^{[3]}$.}
\label{fig4}
\end{figure}

Next, let $n=3$ in {}{T}heorem \ref{thm_nrw}, then the third-order RW of Manakov system is given as
\begin{equation}\label{3_rw}
  q_1^{[3]}=q_2^{[3]}=e^{2{\rm i}t}(\frac{\sqrt{2}}{2}+\frac{F_3}{H_3}).
\end{equation}
Here, $F_3$ and $G_3$ {}{are} two polynomials with degree 12 of $x$ and $t$, which are given in Appendix.
Because $\frac{F_3}{H_3}=-\sqrt{2}$ when $x\rightarrow \infty $ and $t\rightarrow \infty$, the asymptotical
height of  $|q_1^{[3]}|$ and $|q_2^{[3]}|$ is $\frac{\sqrt{2}}{2}$.
Once again $q^{[3]}=\sqrt{2}q_1^{[3]}=\sqrt{2}q_1^{[3]}$ gives the third-order non-symmetrical RW of the NLS,
which has  asymptotical height $1$  when $x\rightarrow \infty $ and $t\rightarrow \infty$.
The profile and the density plot of $q^{[3]}$  are displayed in Fig. \ref{fig4}, which as expected are different from
the symmetrical pattern of the third-order AP \cite{PRE80026601}.

Similar to the case of $q^{[2]}$, it is clearly not symmetric with respect to the spatial co-ordinate, however still the symmetry is conserved with respect to the ($t=0$)-axis. The
co-ordinate of main peak of $|q^{[3]}|$ is indeed not at $(0,\, 0)$. The maximum amplitude occurs at $(0.305,\,0)$ in the ($x,t$)-plane, is of $6.269$. This value is smaller than 7, which characterizes the third-order AP solution \cite{PRE80026601}.
Note also that the latter solution maps six zero amplitude points located on the $x$-axis.

{Since the DT of the Manakov system requires solving a cubic algebraic equation \eqref{eigenfun}, thus, it is possible to derive novel non-trivial solutions for the NLS, when the condition described in Eq. \eqref{transformationnlsandcnls} applies. Indeed, it would be also possible to derive the presented doubly-localized solutions directly from the NLS. This needs further work, which has been started.}  

\subsection{Differences between symmetrical and non-symmetrical RWs}

In the section, we analyze the differences between the derived solution to the family of APs up to third-order. Since the non-symmetric solution of Manakov system can generate the solutions of the NLS equation through the simple scaling transformation $q^{[n]}=\sqrt{2} q_1^{[n]}$, when $c_1=c_2=\frac{\sqrt{2}}{2}$ in Eqs. \eqref{seed}, we just discuss the difference between symmetrical
 and  non-symmetrical solution of the NLS equation. The main characteristic features are summarized in Table 1.

\begin{table}[h]
\centering
\begin{tabular}{|c|c|c|c|c|c|}
\hline
$q_s^{[j]}\&q^{[j]}$     & Symmetric axis        &No. zeros          & Amplitude    &Denom. degree\\\hline
$q^{[1]}$    & $x=\frac{1}{2},\,t=0$      & $2$               &$3$                 &$2$ \\
$q_s^{[1]}$      & $x=0,\,t=0$            & $2$               &$3$                 &$2$ \\\hline
$q^{[2]}$    & $x=0.372,\,t=0$            & $4$               &$4.695$             &$6$ \\
$q_s^{[2]}$      & $x=0,\,t=0$            & $4$               &$5$                 &$6$ \\\hline
$q^{[3]}$    & $x=0.305,\,t=0$            & $6$               &$6.269$             & $12$ \\
$q_s^{[3]}$      & $x=0,\,t=0$            & $6$               &$7$                 & $12$  \\
\hline
\end{tabular}\\
\caption{ The difference between symmetric $q_s^{[j]}$ and non-symmetric $q^{[j]}$ RWs of the NLS equation. Here, \textit{No. zeros} denotes the number of zero amplitudes in $q_s^{[j]}\&q^{[j]}$ along $x$-axis, 
whereas \textit{Denom. degree} labels the degree of the polynomials of $x$ and $t$
 in the denominators of $q_s^{[j]}\&q^{[j]}$. Furthermore, $q_s^{[j]}$ and $q^{[j]}$ represent the symmetric and non-symmetric solution, respectively. The symmetrical fundamental patterns $q_s^{[j]}$ have a fixed asymptotical plane
 with a scaled height of $1$, as constructed in \cite{PRE80026601,PRE87052914}. }
\end{table}

As already mentioned, we can conclude that the solutions $q^{[j]}(j=2,3)$ (the non-symmetric solution) are just symmetrical  with respect to
the $t$-axis and not symmetrical with respect to the $x$-axis. Furthermore, the amplitude of the non-symmetric solution
 is slightly smaller than the corresponding symmetric AP case. Besides, zero amplitudes in $|q^{[j]}(x,0)(j=2,3)|$  are not  distributed along $x$-axis as in the case of symmetrical RWs\cite{PRE80026601}. The number of the zero amplitudes of the RWs $q^{[n]}$  is increasing by increasing the order $n$. Therefore, it is reasonable to infer that the non-symmetric feature becomes more and more remarkable and noticeable, thus, such RW solutions deserve further studies.

\section{Numerical validation and laboratory water wave experiments}

This section is dedicated to the evolution of the derived second-order non-symmetrical RW $q^{[2]}$ in dimensional physical units and to laboratory experiments, which have been conducted in order to observe the latter NLS solution. For that purpose the solution has to be written first in dimensional units, in order to satisfy the deep-water NLS \cite{Zakharov}:
\begin{equation}\label{nls}
i\left(\frac{\displaystyle\partial \psi}{\displaystyle \partial
\tau}+\frac{\displaystyle\omega_0}{\displaystyle2k_0}\frac{\displaystyle\partial \psi}{\displaystyle \partial
\xi}\right)-\frac{\displaystyle\omega_0}{\displaystyle
8k_0^2}\frac{\displaystyle\partial^2 \psi}{\displaystyle \partial
\xi^2}-\frac{\displaystyle\omega_0 k_0^2}{\displaystyle
2}\left|\psi\right|^2\psi=0,
\end{equation}
while the wave frequency $\omega_0$ and the wave number $k_0$ are connected through the dispersion relation $\omega_0=\sqrt{gk_0}$, where $g$ labels the gravitational acceleration. Then, the spatio-temporal evolution of the free surface elevation $\eta(\xi,\tau)$ is approximated to first-order in steepness:
\begin{equation}
\eta(\xi,\tau)=\operatorname{Re}\left(\psi(\xi,\tau)\cdot\exp\left[i\left(k_0\xi-\omega_0\tau\right)\right]\right).
\label{se}
\end{equation}
The amplitude of the carrier $a_0$ and its steepness $\varepsilon_0=a_0k_0$ have been chosen to be of 0.003 m and 0.05, respectively. Therefore, the wave frequency is $\omega_0=12.78\textnormal{ rad}\cdot\textnormal{s}^{-1}$. These values have been chosen in order to satisfy the deep-water gravity waves conditions and to avoid wave breaking, which would significantly violate the evolution of the wave field, as predicted and approximated by the NLS.

\subsection{Numerical MNLS simulations}

\begin{figure}[h]
\centering
\includegraphics[width=10cm]{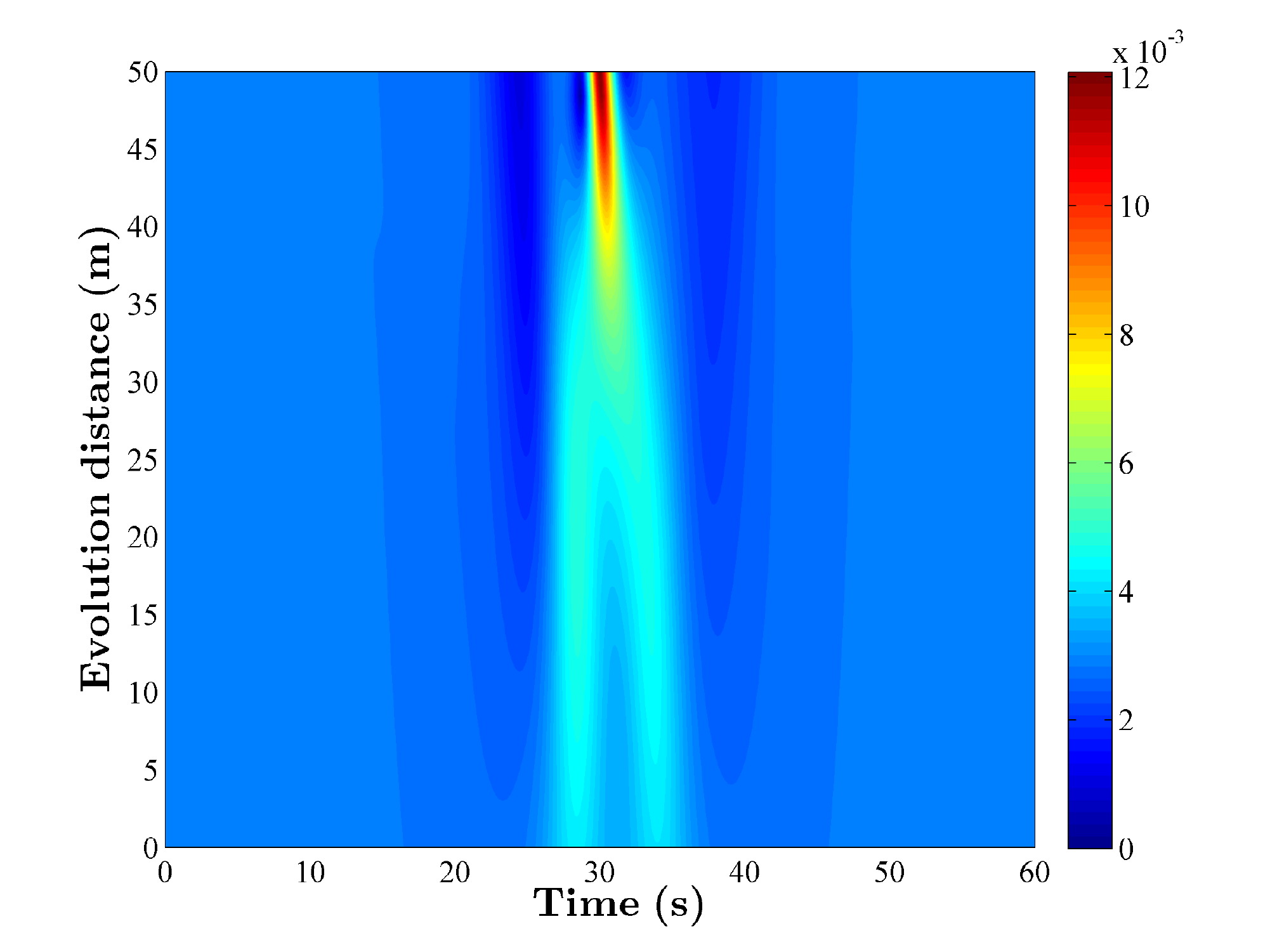}
\caption{MNLS simulations of the dimensional second-order non-symmetric solution $\psi^{[2]}$ for the carrier ampliude $a_0=0.003$ m and $\varepsilon=0.05$, while evolving in space with the group velocity, which is $c_g:=\dfrac{\omega_0}{2k_0}=0.38\textnormal{ m}\cdot\textnormal{s}^{-1}$.}
\label{fig5}
\end{figure}

The first step consists in determining the appropriate initial spatial co-ordinate for the boundary conditions Eq. (\ref{se}), which are needed in order to drive the physical wave maker. Therefore, preliminary numerical simulations based on the modified NLS (MNLS) \cite{Dysthe}, known to describe more accurately the evolution of nonlinear wave packets compared with the NLS since it takes into account higher-order dispersion as well as the mean flow of the wave field, have been performed for the dimensional non-symmetrical second-order $\psi^{[2]}$. Note that as described earlier and differently than the second-order AP solution the maximal amplification of the corresponding field is not at $\xi=0$. Results of the simulations, showing the last stage of envelope propagation over 50 m, before reaching its maximal amplification and for the initial spatial co-ordinate $\xi=-45$ m, are depicted in Fig. \ref{fig5}.

Results in Fig. \ref{fig5} show the evolution of the $\psi^{[2]}$-wave field, which initial perturbation consists of two small envelope modulations, as described by NLS boundary conditions. During this propagation with the group velocity, a strong maximal localization is formed after 50 m and is the result of the nonlinear interaction of the two initially small envelopes.

\subsection{Wave flume experiments}

\begin{figure}[h]
\centering
\includegraphics[width=10cm]{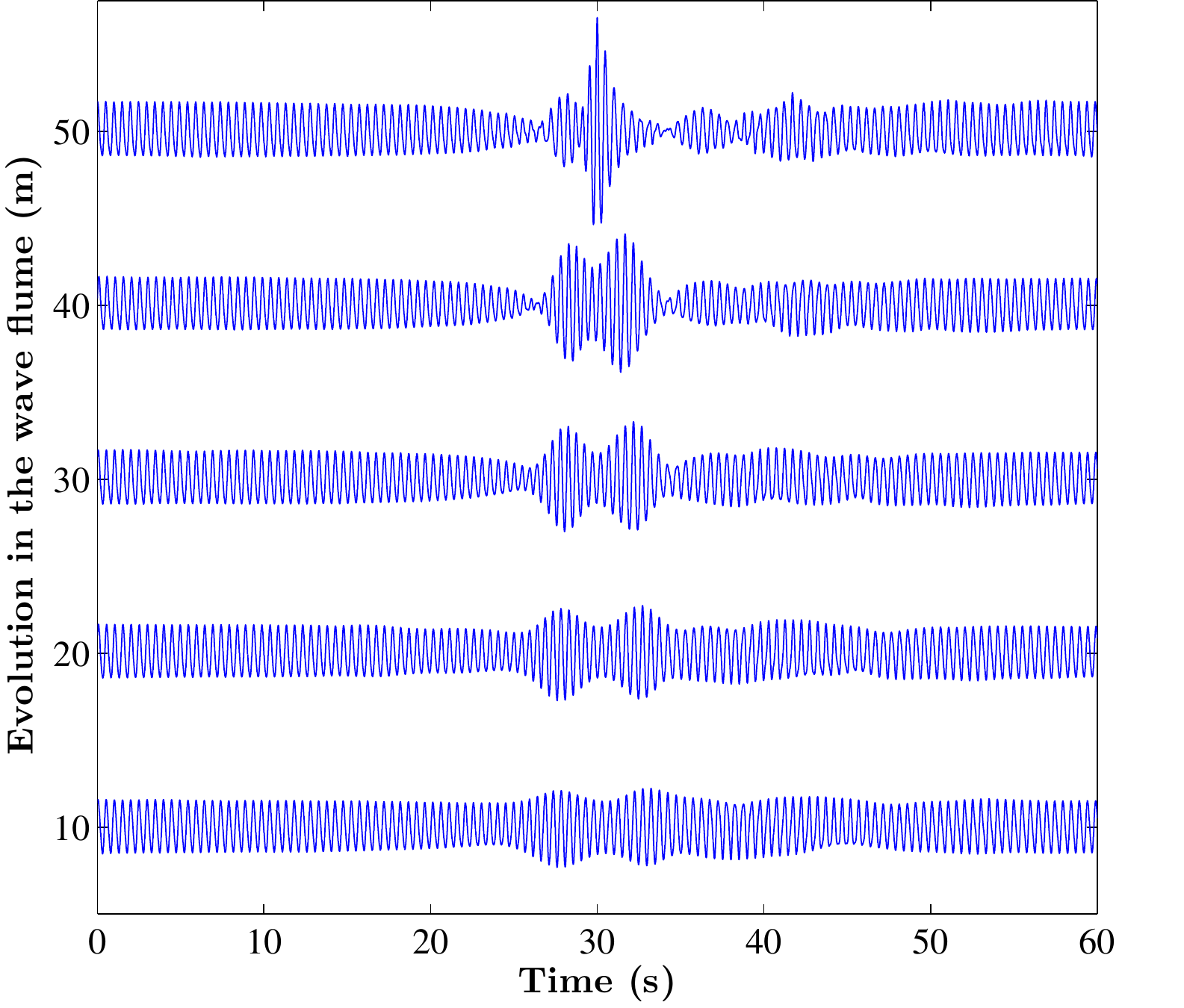}
\caption{Evolution of the second-order non-symmetric solution $\psi^{[2]}$ in a water wave tank for $a_0=0.003$ m and $\varepsilon=0.05$. The five measurements are aligned by the value of the group velocity.}
\label{fig6}
\end{figure}

As next, we will trigger the same evolution dynamics in a laboratory wave flume. Experiments have been conducted in the same facility, as described in detail in \cite{PRX2011015}. The carrier parameters as well as initial conditions have been chosen to be the same as for the numerical simulations of the evolution dynamics. The wave gauge is placed 10 m from the wave maker, which consists of a single flap, controlled by a hydraulic cylinder. The from the gauge measured wave profiles are then reemitted to the flap in order to overcome the length restriction of the wave flume facility and to increase its length, as described in \cite{Chabchoub4}. Therefore, four iteration loops have to be conducted to satisfy a propagation distance of 50 m. The experimental results for the same initial condition as for the numerical simulations are shown in Fig. \ref{fig6}.

\begin{figure}[h]
\centering
\includegraphics[width=10cm]{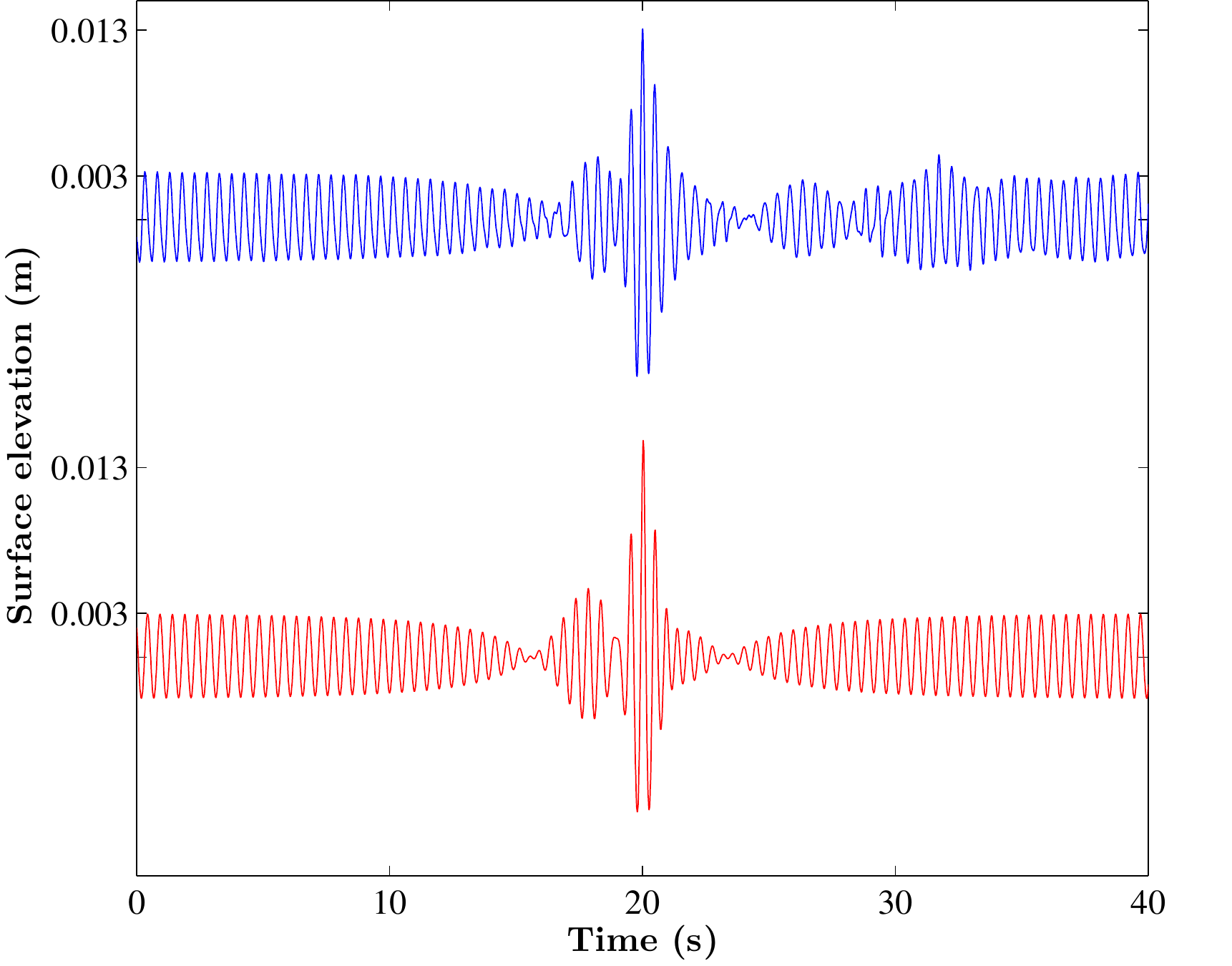}
\caption{Comparison of the maximal measured wave profiles with respect to the theoretical NLS prediction to second-order in steepness.}
\label{fig7}
\end{figure}

As for the doubly-localized second-order AP solution \cite{PRX2011015} and as already provided by the MNLS simulations, the characteristic amplification of the wave field is due to the overlapping two envelope modulations. However, the difference with respect to the symmetric AP solution is that this nonlinear envelope interaction is non-symmetric already at the initial stage of the small amplitude modulations, as can be noticed in Fig. \ref{fig6}. This asymmetrical interaction persists for the whole propagation distance along the flume. Indeed, as predicted by the MNLS simulations, the maximal amplitude amplification is reached after 50 m of wave propagation. Next,
 we compare the maximal wave profile with the NLS prediction. Fig. \ref{fig7} shows the results.

Clearly, the agreement is remarkable. The observed and theoretical wave profiles, shown in Fig. \ref{fig7} are \textit{almost} identical with a noticeable asymmetry compared to the maximal wave, related to this particular solution. Furthermore, the maximal characteristic amplitude amplification of 4.6 is reached. These experimental observations prove that higher-order nonlinear envelopes, generating non-symmetric RWs, can be described by weakly nonlinear evolution equations and can be observed in nonlinear dispersive media, governed by the NLS. The derived non-symmetric solutions naturally have different physical and hydrodynamic properties in terms of surface wave profiles as well as flow field variations, compared with the symmetric AP solutions. This should be analyzed and investigated in more detail numerically and experimentally in order to quantify these physical differences. Furthermore, the results may motivate similar analytical, numerical and experimental studies in other media, governed by the NLS, such as in optics or in plasma.

\section{Conclusion}

To summarize, we presented the determinant expressions of the $n$th-order RWs of the Manakov system in {}{T}heorem 1 and {T}heorem 2,
and gave the explicit expressions for the first-, second- and third-order RW solutions in a particular case such that $q_1^{[j]}=q_2^{[j]}(j=1,2,3)$. As special case, these three solutions determine non-symmetrical RWs $q^{[j]}(j=1,2,3)$ with respect to the spatial co-ordinate under the fundamental framework of the NLS. Fundamental characteristics of these doubly-localized NLS solutions as well as the differences to the corresponding symmetrical AP case, have been discussed in detail. Furthermore, we confirmed the physical validity of the second-order solution $q^{[2]}$, by performing numerical simulations as well as laboratory experiments. The theoretical, numerical and experimental results are in a very good agreement and confirm the accuracy of weakly nonlinear evolution equations to describe symmetric and non-symmetric strongly localized pattern dynamics. This work may motivate further numerical simulations of the fully nonlinear hydrodynamic evolution equations and further experiments in different nonlinear dispersive media.

\mbox{\vspace{1cm}}


{\bf Acknowledgments}  This work is supported by the NSF of China under Grant No.11271210, the K. C. Wong Magna Fund in Ningbo University and the Natural Science Foundation of Ningbo under Grant No. 2011A610179. J. S. H acknowledges sincerely Prof. A. S. Fokas for arranging the visit to Cambridge University in 2012-2014 and for many useful discussions. 
{A. C. acknowledges partial support from the Australian Research Council (Discovery Projects No. DP1093349) and support from the Isaac Newton Institute for Mathematical Sciences.}


\newpage
{\bf Appendix}
{\small
\begin{align*}
  F_3=&2\,\sqrt{2}\left( -256\,{x}^{12}-6144\,{x}^{10}{t}^{2}-61440\,{x}^{8
}{t}^{4}-327680\,{x}^{6}{t}^{6}-983040\,{x}^{4}{t}^{8}-1572864\,{x}^{2
}{t}^{10}\right.\\
&-1048576\,{t}^{12}+1536\,{x}^{11}+3072\,{\rm i}{x}^{10}t+30720\,{x}
^{9}{t}^{2}+61440\,{\rm i}{x}^{8}{t}^{3}+245760\,{x}^{7}{t}^{4}\\
&+491520\,{\rm i}{x}
^{6}{t}^{5}+983040\,{x}^{5}{t}^{6}+1966080\,{\rm i}{x}^{4}{t}^{7}+1966080\,{
x}^{3}{t}^{8}+3932160\,{\rm i}{x}^{2}{t}^{9}\\
&+1572864\,x{t}^{10}+3145728\,{\rm i}{t
}^{11}-3840\,{x}^{10}-15360\,{\rm i}{x}^{9}t-245760\,{\rm i}{x}^{7}{t}^{3}+368640
\,{x}^{6}{t}^{4}\\
&-1474560\,{\rm i}{t}^{5}{x}^{5}+1966080\,{x}^{4}{t}^{6}-
3932160\,{\rm i}{t}^{7}{x}^{3}+2949120\,{x}^{2}{t}^{8}-3932160\,{\rm i}{t}^{9}x+
5760\,{x}^{9}\\
&+23040\,{\rm i}{x}^{8}t-184320\,{x}^{7}{t}^{2}-2027520\,{x}^{5}
{t}^{4}-737280\,{\rm i}{t}^{5}{x}^{4}-5898240\,{x}^{3}{t}^{6}-4423680\,x{t}^
{8}\\
&+5898240\,{\rm i}{t}^{9}-5760\,{x}^{8}+322560\,{x}^{6}{t}^{2}+921600\,{\rm i}{x
}^{5}{t}^{3}+2764800\,{x}^{4}{t}^{4}+4423680\,{\rm i}{x}^{3}{t}^{5}\\
&-3686400
\,{x}^{2}{t}^{6}+2949120\,{\rm i}x{t}^{7}+10321920\,{t}^{8}+4320\,{x}^{7}-
69120\,{\rm i}t{x}^{6}-155520\,{x}^{5}{t}^{2}\\
&-1843200\,{\rm i}{t}^{3}{x}^{4}-
2188800\,{x}^{3}{t}^{4}+1105920\,{\rm i}{x}^{2}{t}^{5}+8755200\,x{t}^{6}-
11796480\,{\rm i}{t}^{7}-2880\,{x}^{6}\\
&+146880\,{\rm i}{x}^{5}t-259200\,{x}^{4}{t}^
{2}+2188800\,{\rm i}{x}^{3}{t}^{3}+1382400\,{x}^{2}{t}^{4}-5391360\,{\rm i}{t}^{5}
x-92160\,{t}^{6}\\
&+4320\,{x}^{5}-216000\,{\rm i}t{x}^{4}+604800\,{x}^{3}{t}^{2
}-2764800\,{\rm i}{t}^{3}{x}^{2}-691200\,x{t}^{4}-5391360\,{\rm i}{t}^{5}\\
&-16200\,{x}^{4}+172800\,{\rm i}{x}^{3}t-1036800\,{x}^{2}{t}^{2}+1555200\,{\rm i}x{t}^{3}-
3542400\,{t}^{4}+21600\,{x}^{3}\\
&\left.+648000\,x{t}^{2}-432000\,{\rm i}{t}^{3}-
18225\,{x}^{2}-64800\,{\rm i}xt-267300\,{t}^{2}+6075\,x+36450\,{\rm i}t \right),\\
H_3=&512\,{x}^{12}+12288\,{x}^{10}{t}^{2}+122880\,{x}^{8}{t}^{4}+655360\,{x
}^{6}{t}^{6}+1966080\,{x}^{4}{t}^{8}+3145728\,{x}^{2}{t}^{10}\\
&+2097152
\,{t}^{12}-3072\,{x}^{11}-61440\,{x}^{9}{t}^{2}-491520\,{x}^{7}{t}^{4}
-1966080\,{x}^{5}{t}^{6}-3932160\,{x}^{3}{t}^{8}\\
&-3145728\,x{t}^{10}+
9216\,{x}^{10}+92160\,{x}^{8}{t}^{2}+491520\,{x}^{6}{t}^{4}+2949120\,{
x}^{4}{t}^{6}+11796480\,{x}^{2}{t}^{8}\\
&+17301504\,{t}^{10}-19200\,{x}^{
9}+368640\,{x}^{5}{t}^{4}-1966080\,{x}^{3}{t}^{6}-8847360\,x{t}^{8}+
34560\,{x}^{8}\\
&-92160\,{x}^{6}{t}^{2}-1843200\,{x}^{4}{t}^{4}+28016640
\,{x}^{2}{t}^{6}+32440320\,{t}^{8}-54720\,{x}^{7}+34560\,{x}^{5}{t}^{2
}\\
&+4377600\,{x}^{3}{t}^{4}-27832320\,x{t}^{6}+86400\,{x}^{6}+518400\,{x
}^{4}{t}^{2}+5529600\,{x}^{2}{t}^{4}+14929920\,{t}^{6}\\
&-108000\,{x}^{5}
-691200\,{x}^{3}{t}^{2}-9331200\,x{t}^{4}+97200\,{x}^{4}+7430400\,{t}^
{4}-64800\,{x}^{3}+259200\,x{t}^{2}\\
&+36450\,{x}^{2}+145800\,{t}^{2}-
12150\,x+2025.
\end{align*}
}
\end{document}